\documentstyle[12pt,amssymb]{article}

\def\rlx{\relax\leavevmode}
\def\inbar{\vrule height1.5ex width.4pt depth0pt}
\def\IZ{\rlx\hbox{\small \sf Z\kern-.4em Z}}
\def\IR{\rlx\hbox{\rm I\kern-.18em R}}
\def\ID{\rlx\hbox{\rm I\kern-.18em D}}
\def\IC{\rlx\hbox{\,$\inbar\kern-.3em{\rm C}$}}
\def\IN{\rlx\hbox{\rm I\kern-.18em N}}
\def\one{\hbox{{1}\kern-.25em\hbox{l}}}

\def\JS{J^{\cal S}}
\def\JL{J^{\cal L}}
\def\half{\frac{1}{2}}
\def\ra{\rightarrow}
\def\beq{\begin{equation}}
\def\eeq{\end{equation}}
\def\bea{\begin{eqnarray}}
\def\eea{\end{eqnarray}}
\def\ber{\begin{array}}
\def\eer{\end{array}}

\def\eps{\varepsilon}
\def\ie{{\it i.e.\ }}
\def\adag{a^\dagger}
\def\bdag{b^\dagger}
\def\half{\frac{1}{2}}
\def\M{{\cal M}}
\newcommand{\bdtxt}[1]{\mbox{\boldmath$#1$}}

\newcommand{\smbox}[1]{{\mbox {\scriptsize #1}}}
\def\gbl{\left[\!\left[}
\def\gbr{\right]\!\right]}
\def\gcl{\left\{ \! \left[}
\def\gcr{\right] \!\right\}}
\def\Ng{N_{\smbox{gh}}}
\newcommand{\sqbr}[3]{\left[\stackrel{#1#2}{\smbox{{\it #3}}}\right]}
\def\Hf{H_{eff}}
\newcommand{\eqn}[1]{(\ref{#1})}

\newcommand{\app}[1]{
      \addtocounter{section}{1}
      \noindent
      {\Large\bf Appendix \thesection:\, #1}
      \setcounter{equation}{0}
   }

\begin{document}

\begin{titlepage}
July 1999 \hfill{UTAS-PHYS-99-11}\\
\vskip 1.6in
\begin{center}
{\Large {\bf Covariant spinor representation of
$\bdtxt{iosp(d,2/2)}$  }}\\[5pt]
{\Large  {\bf and }}\\[5pt]
{\Large {\bf quantization of the
spinning relativistic particle}}
\end{center}

\normalsize
\vskip .4in

\begin{center}
P. D. Jarvis,  \hspace{3pt} S. P. Corney  \hspace{3pt}
and \hspace{3pt} I Tsohantjis
\par \vskip .1in \noindent
{\it School of Mathematics and Physics, University of Tasmania}\\
{\it GPO Box 252-21, Hobart Tas 7001, Australia }\\

\end{center}
\par \vskip .3in

\begin{center}
{\Large {\bf Abstract}}\\
\end{center}

\vspace{1cm}

A covariant spinor representation of $iosp(d,2/2)$ is constructed
for the quantization of the
spinning relativistic particle. It is found that, with appropriately
defined
wavefunctions, this representation
can be identified with the state space arising from the canonical
extended BFV-BRST quantization of the
spinning particle with
admissible gauge fixing conditions after a contraction procedure. For
this model, the cohomological
determination of
physical states can thus be obtained purely from the representation
theory of
the $iosp(d,2/2)$ algebra.

\end{titlepage}

\section{Introduction and Main Results}

In recent work the question of the `super-algebraisation'
of the Hamiltonian BRST-BFV extended phase space quantisation for gauge systems
has been examined\cite{scalar,wigsym,clausthal,d21agrp22}.
Specifically, following earlier indications along these lines in the
literature\cite{casalbuoni},
it has been claimed that there are natural spacetime
`quantization superalgebras' which possess representations precisely
mirroring the BRST-BFV construction in certain cases, namely
relativistic particle systems and generalisations thereof, for which the
relevant
spacetime supersymmetries are the superconformal algebra
$osp(d,2/2)$ and its inhomogeneous extension
\cite{scalar,wigsym,clausthal}, or the family
$D(2,1;\alpha)$
of exceptional superalgebras, in the $1+1$-dimensional case\cite{d21agrp22}.

In a previous work on the scalar particle \cite{scalar} the supersymmetry was realised using the  method of produced superalgebra representations. In the present work this programme is continued with the examination of the covariant BRST-BFV quantisation of the spinning particle model via a spin representation of $iosp(d,2/2)$, and the sharpening of previous work via a covariant tensor notation for this extended spacetime supersymmetry. Our specific results, to be elaborated in the remainder of the paper, are as follows.
In Section \ref{iosp}, a space of covariant spinor superfields
carrying an
appropriate spin representation of $iosp(d,2/2)$ is introduced, and
its structure studied. The generators $J_{MN}$ of $osp(d,2/2)$
have orbital and spin components, associated respectively with
standard configuration space coordinates and differentials $X^{M},
P_{M}=\partial_{M}$, and an extended (graded) Clifford algebra with
generators $\Gamma_{N} $ entailing both fermionic and bosonic
oscillators. The
mass shell condition $P \cdot P - \M^{2} = 0$ factorizes,
allowing the Dirac condition $ \Gamma \cdot \partial - {\cal M} =0$ to
be covariantly imposed at the $d+2/2$ - dimensional level, effecting
a decomposition of the representation space. At the same time, the
Dirac wavefunctions split into upper and lower components, so that the
$iosp(d,2/2)$ algebra is effectively realised
on $2^{d/2}$-dimensional Dirac spinors (over $x^{M}$, and subject to a
certain differential constraint on $P_{-}$, deriving from the mass
shell condition). Full technical details of the construction are
given in Section \ref{iosp}.

In Section \ref{physstates}  a `BRST operator' $\Omega$ is named as
one of the nilpotent
odd generators of the homogeneous superalgebra (a `super-boost'
acting between
fermionic and light cone directions), relative to a choice of `ghost
number' operator within the $sp(2)$ sector. Correspondingly a `gauge
fixing
fermion' ${\cal F}$ of opposite ghost number is identified
(a `supertranslation' generator), and physical Hamiltonian $H = \{{\cal F}, \Omega \}$. Finally, the cohomology of $\Omega$ is constructed
at arbitrary ghost number.
It is found that the `physical states' thus defined are precisely
those wavefunctions which obey the conventional $(d-1)+1$-dimensional
Dirac equation, and moreover which have a fixed degree of homogeneity in
the light cone coordinate $p_{+}$. Given that the $P_{-}$ constraint
already dictates the evolution of the Dirac spinors in the light
cone time $x^- = \eta^{-+}x_{+}$, the analysis thus reveals that this
`superalgebraisation' of the BRST-BFV quantisation yields
the correct spin-$\frac 12$ irreducible representation of the
Poincar\'{e} algebra in $(d-1)+1$-dimensions, as carried on the space of
covariant solutions of the massive Dirac equation.

As this construction has been obtained purely algebraically, without
the use of a physical model, it is finally the task of Section
\ref{brstbfv} to establish that the standard Hamiltonian BRST-BFV {\it Ansatz}, applied to the spinning particle model\cite{brink,casalbuoni,brinkschwarz,brinkgreen}, does indeed give rise to an identical state space structure. The only proviso on this statement turns out to be that the model's extended phase space should formally be modified by a contraction, or `$\beta$ - limit' \cite{govbk} in order to identify the appropriate sector of the full phase space (for details see Section \ref{betalimit}).

In conclusion, the import of our programme,
exemplified by the present case study, is an approach to covariant quantisation
of models with gauge symmetries via a cohomological
realisation of the appropriate space of irreducible representations
of physical states (in the case of particle models on flat spacetime, the Poincar\'{e} algebra) through the construction of the correct BRST complex
(in the present cases, as realised within the covariant
representations of the `quantisation superalgebras' $osp(d,2/2)$ and
generalisations). Further concluding remarks, and prospects for future
work, are given in
Section \ref{conc}.

\section{Covariant representations of $iosp(d,2/2)$}
\label{iosp}
\subsection{Introduction and notation}

The $iosp(d,2/2)$ superalgebra is a generalisation of $iso(d,2)$. The
supermetric $\eta_{MN}$ we shall use throughout is made up of three parts; the first
has block diagonal form with the entries being the Minkowski metric tensor of
$so(d,1)$ with $-1$ occurring $d$ times,
\[
\eta_{\mu\nu} = diag(1,-1,-1,\ldots,-1).
\]
The second part is off diagonal and can be written
\[
\eta_{ab} = \left( \begin{array}{cc} 0&1 \\ 1&0 \end{array} \right),
\]
where $a,b = \pm$, reflecting a choice of light cone coordinates in
two additional bosonic dimensions, one spacelike and one timelike.
The final part corresponds to the Grassmann odd components and is the
symplectic metric tensor
\[
\eta_{\alpha\beta} = \eps_{\alpha\beta} = \left( \begin{array}{cc}
0&1 \\ -1&0 \end{array} \right).
\]
Here Greek indices $\alpha,\beta,\ldots$ take values $1,2$, whilst
$\lambda,\mu,\nu,\ldots$ take values in the range $0,\ldots,d-1$, whilst Latin indicies $a,b,c,\ldots$ range over $\mu,\nu,\ldots, +, -$. The
indices $M,N,\ldots$ cover all values, and thus run over
$0,\ldots,d-1,+,-,1,2$. We also define graded commutator brackets as
\bea
\gbl A_M,B_M \gbr &=& A_M B_N - [MN] B_N A_M, \nonumber \\
\gcl A_M,B_M \gcr &=& A_M B_N + [MN] B_N A_M, \nonumber
\eea
where the convention for the $[MN]$ sign factor is $[MN] = (-1)^{mn}$ (extended
to $\sqbr{M}{N}{P} = (-1)^{mn+np}$ as necessary). Note that the
grading factors are $m,n = 0$ for Minkowski and light cone indices
$\mu,\nu, \ldots, \pm$
and $m,n = 1$ for symplectic indices $\alpha, \beta, \ldots$. With these index conventions the metric thus obeys $\eta_{MN} = [MN] \eta_{NM}$.

We define $J_{MN}=-[MN]J_{NM}$ as the generators of the $osp(d,2/2)$
superalgebra, with commutation relations as follows \cite{pdjhsg}
\beq
\gbl J_{MN},J_{PQ} \gbr = -\eta_{NQ}J_{MP} + [NP] \eta_{NP} J_{MQ} - [MN] [MP] \eta_{MP} J_{NQ} + [PQ][MN][MQ] \eta_{MQ} J_{NP}.
\label{eq:homopart}
\eeq
The homogeneous
even subalgebra is $so(d,2)\oplus sp(2,\IR)$ with $so(d,2)$ generated
by $J_{\mu\nu} = - J_{\nu\mu}$, and $sp(2,\IR)$ by
$J_{\alpha\beta} = J_{\beta\alpha}$. For clarity,
we set $J_{\alpha\beta} \equiv K_{\alpha\beta} = K_{\beta\alpha}$.
Likewise, the odd
generators will be denoted $J_{\mu\alpha} \equiv L_{\mu\alpha}$ or
$J_{\alpha\pm} \equiv L_{\alpha\pm}$. The
inhomogeneous part $i(d,2/2)$ consists of additional (super)translation generators $P_{M}$ satisfying
\[
\gbl J_{MN},P_{L} \gbr = \eta_{LN} P_M - [MN]\eta_{LM} P_N, \nonumber 
\label{eq:inhomopart}
\]
The $d+2$ even translations are
$P_\mu, P_{\pm}$ acting in the $(d,2)$ pseudo-Euclidean space, and
the two odd nilpotent supertranslations are $P_{\alpha} \equiv Q_\alpha$.

We consider a class of covariant spinor superfield representations of
$iosp(d,2/2)$ (compare \cite{scalar,hartley}) acting on
suitable spinor wavefunctions $\Psi(x^{M})$ over $d+2/2$-dimensional
superspace\footnote{${\cal S}$ denotes the superfields over
$d+2/2$-dimensional superspace $(x^{M}) = (x^{\mu},
x^{\pm},\theta^{\alpha})$, while ${\cal B}\otimes{\cal F}$
carries the graded Clifford algebra (see below)}, $({\cal B}\otimes{\cal F}\otimes{\cal S})$.

The $osp(d,2/2)$ generators can be more explicitly written
\beq
J_{MN} = \JL_{MN} + \JS_{MN},
\label{eq:totalj}
\eeq
where the orbital part is defined
\beq
\JL_{MN} = X_M \partial_N - [MN] X_N \partial_M,
\eeq
with 
\[
\partial_N = P_N = \frac{\partial}{\partial X^N} = \left( \frac{\partial}{\partial X^\mu},\frac{\partial}{\partial X^\pm},\frac{\partial}{\partial X^\alpha}\right).
\]
The spin part of \eqn{eq:totalj} is
\beq
\JS_{MN}= \frac{1}{4}\gbl \Gamma_M,\Gamma_N \gbr,
\eeq
where $\Gamma_M, \Gamma_N$ are
generalised Dirac matrices. Of course both $\JL$ and $\JS$ fulfill the
$osp(d,2/2)$ algebra.

The graded Clifford algebra with generators $\Gamma_N$,  acting on
the space\footnote{${\cal B}$ is the bosonic part which carries the representation of
$\zeta_\alpha$ (see below), while the fermionic ${\cal F}$ carries the
representation of the Dirac
algebra $\gamma_\mu, \gamma_\pm$ and $\gamma_5$} $({\cal B}\otimes{\cal F})$, is
defined through
\beq
\gcl \Gamma_M,\Gamma_N \gcr = \Gamma_M \Gamma_N + [MN] \Gamma_N
\Gamma_M = 2 [MN] \eta_{MN},
\eeq
( if $M \neq N$ then we can write $\Gamma_M \Gamma_N = - [MN]
\Gamma_N \Gamma_M$). Writing the $\Gamma$ in compact form as $\Gamma_M =
\left( \Gamma_\mu, \Gamma_+, \Gamma_-, \Gamma_\alpha \right)^T,$
we have
\bea
\Gamma_\mu & = & 1\otimes \hat{\gamma}_\mu \otimes 1, \nonumber \\
\Gamma_\pm & = & 1 \otimes \hat{\gamma}_\pm \otimes 1, \\
\Gamma_\alpha & = & \zeta_\alpha \otimes \hat{\gamma}_5 \otimes
(-1)^z, \nonumber
\eea
where
\beq
\begin{array}{rclrcl}
\hat{\gamma}_{\mu} &=& \left( \begin{array}{cc}
			\gamma_\mu & 0 \\ 0 & - \gamma_\mu
			\end{array} \right), &
\hat{\gamma}_{+} &=& \sqrt{2}\left( \begin{array}{cc}
			0 & 0 \\ 1 & 0
			\end{array} \right), \\
\hat{\gamma}_{-} &=& \sqrt{2}\left( \begin{array}{cc}
			0 & 1 \\ 0 & 0
			\end{array} \right), &
\hat{\gamma}_{5} &=& \left( \begin{array}{cc}
			\gamma_5 & 0 \\ 0 & -\gamma_5
			\end{array} \right), \\
\end{array}
\eeq
$(-1)^z$ is the parity operator for the $\theta$s and is such that $(-1)^z\theta_\alpha = -\theta_\alpha$, and $(-1)^z \theta_\alpha \theta_\beta =\theta_\alpha \theta_\beta$, and is defined such that $z \equiv \theta^\alpha \partial_\alpha$, $\gamma_5$ is defined such that $\gamma_5^2 = \kappa_5 (=\pm 1)$.

The definition of $\JS_{\alpha\beta}$ leads to
\bea
4 \JS_{\alpha\beta} &=& \gbl \Gamma_\alpha, \Gamma_\beta \gbr =
\zeta_\alpha \gamma_5 \zeta_\beta \gamma_5 (-1)^{2z} \left(
\begin{array}{cc}1 & 0 \\ 0&1 \end{array} \right) + \zeta_\beta
\gamma_5 \zeta_\alpha \gamma_5 (-1)^{2z} \left( \begin{array}{cc}1 &
0 \\ 0&1 \end{array} \right), \nonumber \\
&=& \kappa_5 \{ \zeta_\alpha,\zeta_\beta \}
\left( \begin{array}{cc} 1 & 0 \\ 0 & 1 \end{array} \right);
\eea
moreoever $\gcl \Gamma_\alpha, \Gamma_\beta \gcr = - 2
\eps_{\alpha\beta}$
and so we take
\beq
[\zeta_\alpha,\zeta_\beta] = -2 \kappa_5\eps_{\alpha\beta}.
\eeq
In Appendix A we provide a realisation of $\zeta_\alpha$ (with $\kappa_5 = 1$)
in terms of a pair of bosonic oscillators with indefinite metric.

From \eqn{eq:homopart}, \eqn{eq:inhomopart} it is easy to establish the invariance
of the square of the momentum operator, namely
\bea
\gbl J_{MN},P^R P_R \gbr &=& \left(\delta^R_N P_M - [MN]
\delta^R_M P_N \right)  + P^R  \sqbr{M}{R}{N}\left(\eta_{RN}P_M - [MN] \eta_{RM} P_N \right), \nonumber \\
& = & 0,
\eea
Thus the second order Casimir is
\beq
\label{massoperator}
P^M P_M \equiv P^{\mu}P_{\mu}+ Q^{\alpha}Q_{\alpha}
\eeq

Similarly we get the required generalisation of the Pauli-Lubanski operator \\ $[J_{MN}, W^{ABC}W_{CBA}] =
0$, providing a fourth order Casimir operator.
For any vector operator $V_A$ we have
\beq
[J_{MN},V_A] = \eta_{AM}V_N - [MN] \eta_{AN} V_M,
\eeq
and similarly for any tensor operator $V_{AB}, V_{ABC}$, for example
\bea
[J_{MN},V_{ABC}] &=& \eta_{AM}V_{NBC} - [MN] \eta_{AN} V_{MBC} +
\sqbr{M}{A}{N} \left(\eta_{BM} V_{ANC} - [MN] \eta_{BN}V_{AMC}
\right) \nonumber \\
& & + \sqbr{M}{A}{N}\sqbr{M}{B}{N} \left( \eta_{CM} V_{ABN} - [MN]
\eta_{CN} V_{ABM} \right).
\eea
From this we can calculate
\beq
[J_{MN},V^{ABC}V_{CBA}] = \eta^{AD} \eta^{BE} \eta^{CF} [J_{MN},V_{DEF}V_{CBA}] =0.
\eeq
If we define $V_{ABC} = W_{ABC} = P_A J_{BC} + \sqbr{B}{C}{A} P_C J_{AB} + \sqbr{B}{A}{C} P_B J_{CA}$ we get the required identity.

\subsection{Dirac condition and reduced realisation of $iosp(d,2/2)$
superalgebra}
\label{sec:realis}

In order to project out irreducible representations of the full
superalgebra, we require the mass shell condition (Klein-Gordon equation): in representation terms, a requirement for reducibility of the $iosp(d,2/2)$ representation.
\beq
\left(P^M \eta_{MN} P^N - \M^2 \right)\Psi  =0.
\label{eq:dirac}
\eeq
However, using the Clifford algebra just defined we have
$\gbl J_{MN},\Gamma^L P_L\gbr = 0$ and so we can covariantly impose
the stronger Dirac condition,
\bea
0&=& \left(P^M \eta_{MN} P^N - \M^2 \right), \nonumber
\\
&=&\left(P^M\Gamma_M + \M\right)\left(P^M\Gamma_M - \M\right), \nonumber
\eea
\ie, taking for example the positive root,
\beq
\left( P^M\Gamma_M - \M \right) \Psi = 0.
\label{eq:const}
\eeq

We now construct the explicit forms of the generators $J_{MN}$ of
$iosp(d,2/2)$ within this decomposition of the full space.
Expanding the sum in \eqn{eq:const}  gives
\[
\left( \Gamma^\mu P_\mu + \Gamma^+ P_+ + \Gamma^- P_- + \Gamma^\alpha
Q_\alpha - \M \right) \Psi = 0,
\]
or, in the explicit form, writing $\Psi$ as a two component array
$\Psi = \left( \begin{array}{c} \psi \\ \sqrt{2} \phi \end{array}
\right)$, gives
\[
\left( \begin{array}{cc} \gamma^\mu P_\mu + \zeta^\alpha (-1)^z
\gamma_5 Q_\alpha- \M & \sqrt{2} P_+ \\ \sqrt{2} P_- &
-\left[\gamma^\mu P_\mu + \zeta^\alpha (-1)^z \gamma_5 Q_\alpha + \M
\right]\end{array} \right)
\left( \begin{array}{c}\psi \\ \sqrt{2} \phi \end{array} \right) = 0.
\]
From this, we get the rather useful expression
\beq
\phi = -\frac{1}{2 P_+} \left( \gamma^\mu P_\mu + \zeta^\alpha (-1)^z
\gamma_5 Q_\alpha - \M \right) \psi,
\label{eq:phi}
\eeq
and so $P_-$ can be written
\beq
P_- \psi = - \frac{1}{2 P_+} (\gamma^\mu P_\mu + \zeta^\alpha (-1)^z
\gamma_5 Q_\alpha + \M)(\gamma^\nu P_\nu + \zeta^\beta (-1)^z
\gamma_5 Q_\beta - \M ) \psi.
\eeq
Simplifying this equation yields
\bea
P_- \psi &=&  - \frac{1}{2 P_+} \left(P^2 + Q_\alpha
\eps^{\beta\alpha}Q_\beta \kappa_5 - \M^2\right) \psi, \nonumber \\
&=& - \frac{1}{2 P_+} \left(P^2 + Q^\alpha Q_\alpha\kappa_5 -
\M^2\right) \psi,
\label{eq:KGeqn}
\eea
which we shall later use as the Hamiltonian. This equation is basically the
Klein-Gordon equation (see equation \eqn{eq:dirac}) of the BFV quantised spinning relativistic particle model which will carry the representation.

We are now in a position to explicitly determine the generators of
$osp(d,2/2)$. We show below the process for calculating
$J_{\alpha-}$, and then state without proof all other terms, with the
understanding that the same process is repeated for each: \newline
We have $J_{\alpha-} = \JL_{\alpha-} + \JS_{\alpha-}$, where
$\JL_{\alpha-} = X_\alpha P_- - X_- P_\alpha$ and
\bea
\JS_{\alpha-} &=& \frac{1}{4} \gbl \Gamma_\alpha,\Gamma_- \gbr =
\half \Gamma_\alpha \Gamma_-, \nonumber\\
 &=& \half \left( \begin{array}{cc} \zeta_\alpha \gamma_5 (-1)^z & 0
\\
		0 & -\zeta_\alpha \gamma_5 (-1)^z \end{array} \right)
\left( \begin{array}{cc} 0 & \sqrt{2} \\ 0 & 0 \end{array} \right),
\nonumber \\
& = & \left( \begin{array}{cc} 0 & \frac{1}{\sqrt{2}} \zeta_\alpha
\gamma_5 (-1)^z \\ 0 & 0 \end{array} \right),
\eea
and so
\bea
J_{\alpha-} \Psi & = & \left( \begin{array}{cc}
X_\alpha P_- - X_- P_\alpha & \frac{1}{\sqrt{2}}\zeta_\alpha \gamma_5
(-1)^z \\
0 & -X_\alpha P_- + X_- P_\alpha \end{array} \right)
\left( \begin{array}{c} \psi \\ \sqrt{2} \phi \end{array} \right),
\nonumber\\
& = &\left( \begin{array}{cc} (X_\alpha P_- - X_- P_\alpha)\psi +
\zeta_\alpha \gamma_5 (-1)^z \phi \\ -(X_\alpha P_- - X_-
P_\alpha)\sqrt{2}\phi \end{array} \right). \nonumber
\eea
Substituting in from equation \eqn{eq:phi} gives
\bea
J_{\alpha-} &=& X_\alpha P_- - X_- P_\alpha \nonumber \\
& & \hspace*{10mm} - \frac{\zeta_\alpha}{2 P_+} \left(\gamma_5 (-1)^z
\gamma \cdot P + \gamma_5(-1)^z\zeta^\beta(-1)^z\gamma_5 P_\beta -
\gamma_5(-1)^z \M \right), \nonumber\\
&=& X_\alpha P_- - X_- P_\alpha - \zeta_\alpha \zeta_-,
\label{eq:jalpha-}
\eea
where
\beq
\zeta_- = \frac{1}{2 P_+} \left(\gamma_5 (-1)^z \gamma \cdot P +
\zeta^\beta\kappa_5 P_\beta - \gamma_5(-1)^z \M \right).
\label{eq:xi-}
\eeq

The remaining generators are:
\beq
\begin{array}{rclrcl}
J_{\mu-} &=& X_\mu P_- - X_- P_\mu - \zeta_\mu \zeta_-, &
J_{\mu\alpha} &=& X_\mu P_\alpha - X_\alpha P_\mu + \frac{\kappa_5}{2}
\zeta_\mu \zeta_\alpha, \\
J_{\mu\nu} &=& X_\mu P_\nu - X_\nu P_\mu - \frac{\kappa_5}{4}
[\zeta_\mu,\zeta_\nu], &
J_{+\mu} &=& X_+ P_\mu - X_\mu P_+, \\
J_{+\alpha} &=& X_+ P_\alpha - X_\alpha P_+, &
J_{\alpha\beta}& =& X_\alpha P_\beta + X_\beta P_\alpha +
\frac{\kappa_5}{4} \left\{ \zeta_\alpha, \zeta_\beta \right\}, \\
J_{+-} &=& X_- P_+ - X_+ P_- - \half, & &&
\end{array}
\label{eq:js}
\eeq
where we have defined
\beq
\zeta_\mu = \frac{\gamma_\mu\gamma_5(-1)^z}{\kappa_5}.
\label{eq:ximu}
\eeq
From the above definition we can easily show that $[\zeta_\mu,
\zeta_\nu ] =  -\frac{1}{\kappa_5} [\gamma_\mu,\gamma_\nu]$. The
non-zero commutation relations between $\zeta_-$ and
the remaining operators for $iosp(d,2/2)$ can be calculated and are:
\bea
\{ \zeta_\mu, \zeta_- \} & = & \gcl
\frac{\gamma_\mu\gamma_5(-1)^z}{\kappa_5}, \frac{1}{2P_+} \left(
\gamma_5(-1)^z \gamma\cdot P + \kappa_5\zeta^\beta P_\beta - \gamma_5
(-1)^z \M \right) \gcr, \\
&=&\frac{1}{2\kappa_5 P_+} \gcl\gamma_\mu \gamma_5,\gamma_5\gamma^\nu
P_\nu \gcr - \frac{\M}{\kappa_5} \gcl \gamma_\mu\gamma_5,\gamma_5
\gcr, \\
&=& \frac{P_\mu}{P_+},
\eea
and
\beq
\begin{array}{ccc}
\{\gamma_5,\zeta_-\} = -\frac{\kappa_5}{P_+}\M, & \{\zeta_-,\zeta_-\}
= \frac{P_-}{P_+}, & [X_- , \zeta_- ] = -\frac{\zeta_- }{P_+}, \\
\left[ X_\mu , \zeta_- \right] = - \frac{\kappa_5}{2P_+} \zeta_\mu, 
 & \{X_\alpha,\zeta_-\} = -\frac{\kappa_5\zeta_\beta}{2P_+}, &
\{\zeta_-,\zeta_\alpha \} = \frac{\kappa_5 Q_\alpha}{P_+}.
\end{array}
\label{eq:rawmat3}
\eeq

In summary, the realisation of $iosp(d,2/2)$ that we use is
formulated in terms of the operators $X^{\mu}, P_{\mu} =
\frac{\partial}{\partial x^\mu}, \gamma_\mu, \gamma_5$, together with
$X^\alpha = \theta^\alpha, P_\alpha = Q_\alpha =
\frac{\partial}{\partial \vartheta^\alpha}, \zeta_\alpha, \zeta_-$, and $X_+ =
\tau I, P_- = H, P_+, X_-$. The non-zero commutation relations
amongst these variables are
\beq
\begin{array}{ccc}
\left[X_\mu ,P_\nu \right]=-\eta_{\mu\nu}, &
\left\{\theta_\alpha,Q_\beta \right\} = \eps_{\alpha\beta},
&\left[X_-,P_+\right]=1,  \\
\left[X_-,P_-\right] = - P_+^{-1} P_-, &  \left[\theta_\alpha
,P_-\right] =P_+^{-1} Q_\alpha , &
\left[X_\mu,P_-\right]=P_+^{-1}P_\mu,
\end{array}
\label{eq:rawmat1}
\eeq
and
\beq
\begin{array}{cc}
\{ \zeta_\mu, \zeta_\nu \} = - 2\kappa_5 \eta_{\mu\nu}, &
[\zeta_\alpha,\zeta_\beta ] = -2\kappa_5 \eps_{\alpha\beta}.
\end{array}
\label{eq:rawmat2}
\eeq
Note in the above that $X_+$ and $P_-$ are no longer canonically conjugate when acting on the $\psi$ part of the superfield.

We have now calculated the complete set of non-zero commutation
relations between the operators $X^M, P_N, \zeta_M, \gamma_5$ and
have shown that they do indeed provide the correct realisation of
$iosp(d,2/2)$ on the $\psi$ superfields.
Remarkably, {\it precisely} these operators will emerge as the raw
material in the extended BFV-BRST Hamiltonian quantisation of the
relativistic spinning particle model (Section \ref{brstbfv} below).
However, the algebraic setting already provides the means to complete
the cohomological construction of physical states, as we now show.

\section{Physical States}
\label{physstates}

The physical states of a system can be determined by looking at the
action of the BRST operator $\Omega$ and the ghost number operator
$\Ng$ upon arbitrary states $\psi, \psi'$. As is well known\cite{govbk},
the physical states obey the equations
\[
\Omega \psi = 0, \quad \psi \ne \Omega \psi', \mbox{ and } \Ng \psi =
\ell \psi,
\]
for some eigenvalue $\ell$, where $\Omega$ is the BRST operator,
and $\Ng$ is the ghost number. Therefore in order to determine the
physical states we shall fix $\Omega$ and $\Ng$, and determine  their actions
upon an arbitrary spinor-valued superfield $\psi$.

\noindent
Take two $c$-number $sp(2)$ spinors $\eta^\alpha, \; \eta'^\alpha$ with the
following
relations
\bea
\eta^\alpha\eta_\alpha &= 0 =& \eta'^\alpha\eta'_\alpha, \nonumber\\
\eta^\alpha \eta'_\alpha & = 1 = & - \eta'^\alpha \eta_\alpha.
\label{eq:spinors}
\eea
An example of two such spinors is
\beq
\eta = \frac{1}{\sqrt{2}} \left( \begin{array}{c} 1\\ 1 \end{array}
\right), \; \eta' = \frac{1}{\sqrt{2}} \left( \begin{array}{c} -1 \\
1 \end{array} \right).
\eeq
Below in the superfield expansions we use
\beq
\begin{array}{cc}
\theta_\eta = \eta^\alpha \theta_\alpha, & \theta'_{\eta} =
\eta'^\alpha \theta_\alpha, \\
\chi_\eta = \eta^\beta \chi_\beta, & \chi'_\eta = \eta'^\beta \chi_\beta.
\end{array}
\eeq
The first of these pairs of definitions leads to
\[
\frac{\partial}{\partial \theta^\alpha} = \frac{\partial
\theta_\eta}{\partial \theta^\alpha}\frac{\partial}{\partial
\theta_\eta} + \frac{\theta'_{\eta}}{\partial
\theta^\alpha}\frac{\partial}{\partial\theta_{\eta'}} = -
\eta_\alpha\frac{\partial}{\partial\theta_\eta} - \eta'_\alpha
\frac{\partial}{\partial \theta'_{\eta}},
\]
and therefore
\bea
\eta^\alpha \frac{\partial}{\partial \theta^\alpha} = -
\frac{\partial}{\partial \theta'_{\eta}} &\mbox{ and }& \eta'^\alpha
\frac{\partial}{\partial \theta^\alpha} = - \eta'^\alpha \eta_\alpha
\frac{\partial}{\partial \theta_\eta} = \frac{\partial}{\partial
\theta_\eta}.
\eea

Choose the BRST operator\footnote{The corresponding anti-BRST operator is $\bar{\Omega} = \eta'^\alpha L_{\alpha -}$} and gauge fixing operators as 
\beq
\begin{array}{rcl}
\Omega &=& \eta^\alpha L_{\alpha -}, \\
{\cal F} &=& \eta'^\alpha P_\alpha,
\end{array}
\label{eq:BRSTandPsi}
\eeq
and consistently the ghost number operator $\Ng \equiv
\eta^\alpha\eta'^\beta K_{\alpha\beta}$ satisfies
\bea
[ \Ng, \Omega ] =\Omega, & \mbox{and} & [\Ng, {\cal F} ]
= - {\cal F}, \nonumber
\eea
as required.

\subsection{Action of ghost number operator}

Note that in our case
\[
\Ng = \eta^1\eta'^1 K_{11} + \eta^1\eta'^2 K_{12} +\eta^2\eta'^2
K_{22} +\eta^2\eta'^1 K_{21} = - \half (K_{11} - K_{22}).
\]
$K_{\alpha\beta}$ can be written $K_{\alpha\beta} = K^S_{\alpha\beta}
+ K^L_{\alpha\beta}$ where the two parts denote the bosonic and fermionic
(spin and orbital) contributions respectively. This leads to $\Ng$ having
two parts as well;
\beq
K^L_{\alpha\beta} =  \theta_\alpha \frac{\partial}{\partial
\theta^\beta}  + \theta_\beta \frac{\partial}{\partial \theta^\alpha},
\eeq
therefore
\beq
N^L_{\smbox{gh}} = \theta_\eta \frac{\partial}{\partial \theta_\eta}
- \theta'_{\eta} \frac{\partial}{\partial \theta'_{\eta}},
\eeq
and for the bosonic sector
\beq
K^S_{\alpha\beta} = \frac{1}{4} \gbl \Gamma_\alpha, \Gamma_\beta \gbr
= \frac{1}{4} \{\zeta_\alpha,\zeta_\beta \}\kappa_5,
\eeq
therefore
\bea
N^S_{\smbox{gh}} & = & \frac{\kappa_5}{4} \left( (\eta \cdot \zeta)(
\eta'\cdot \zeta) + (\eta'\cdot \zeta)( \eta \cdot \zeta) \right),
\nonumber\\
& = & \frac{\kappa_5}{2}(\eta\cdot\zeta)(\eta'\cdot\zeta) +
\frac{\kappa_5}{4} =
\frac{\kappa_5}{2}(\eta'\cdot\zeta)(\eta\cdot\zeta) -
\frac{\kappa_5}{4}.
\eea

As seen in Appendix A, we can write a series expansion of
an arbitrary spinor superfield $\psi$ over $(x^\mu,x^\pm,\theta^\alpha)$
in an occupation number basis in the indefinite metric space acted on
by $\zeta_{\alpha}$,
\bea
\psi &=& \sum_{m,n=0}^\infty \psi^{(m,n)}|m,n\rangle
= \sum_{m,n=0}^\infty \left(A^{(m,n)} + \theta^\gamma {\chi^{(m,n)}}_\gamma +
\half \theta^2 B^{(m,n)} \right)|m,n\rangle,
\label{eq:psi} \\
&=& A + \theta^C \chi_C + \half \theta^2 B + \ldots. \nonumber
\eea
We can re-write this series expansion with respect to the spinors
\eqn{eq:spinors} as follows
\bea
\theta^\alpha \chi_\alpha & = & \theta^\alpha \delta^\beta_\alpha
\chi_\beta,
 = \theta^\alpha \left( \eta^\beta \eta'_\alpha - \eta_\alpha
\eta'^\beta \right) \chi_\beta,\nonumber  \\
& = & \theta_\eta \chi'_\eta - \theta'_\eta \chi_\eta,
\label{eq:theta}
\eea
and
\bea
\half \theta^2 &=& \half \theta^\alpha
\eps_{\alpha\beta}\theta^\beta,
 =  \half \theta^\alpha \left( \eta_\alpha \eta'_\beta - \eta_\beta
\eta'_\alpha \right) \theta^\beta, \nonumber \\
& = & \half \left( \theta_\eta \theta'_\eta - \theta'_\eta
\theta_\eta \right) = \theta_\eta \theta'_\eta. \label{eq:thetasq}
\eea
Thus using equations \eqn{eq:theta} and \eqn{eq:thetasq},
\beq
\psi ^{(m,n)} = A^{(m,n)}+ \theta_\eta \chi'^{(m,n)}_\eta - \theta'_\eta
\chi^{(m,n)}_\eta + \theta_\eta \theta'_\eta B^{(m,n)}.
\label{eq:psispin}
\eeq
In what follows, the occupation number labels in the bosonic space will
be suppressed, whereas the structure of the explicit superfield
expansion will be needed. Thus for example
$A \equiv \sum_{m,n=0}^\infty A^{(m,n)}$. Note
\beq
\begin{array}{cc}
N^L_{\smbox{gh}} A = 0, & N^L_{\smbox{gh}}(\theta_\eta \chi'_\eta) =
\theta_\eta\chi'_\eta, \\
N^L_{\smbox{gh}}(\theta_\eta\theta'_\eta B) = 0, &
N^L_{\smbox{gh}}(-\theta'_\eta \chi_\eta) = \theta'_\eta\chi_\eta,
\end{array}
\eeq
and so
\bea
\Ng \psi & = & \half \left( \kappa_5(\eta\cdot\zeta)(\eta'\cdot\zeta) +
\frac{\kappa_5}{2}\right)A + \half\theta_\eta\left(
\kappa_5(\eta\cdot\zeta)(\eta'\cdot\zeta) + \frac{\kappa_5+4}{2} \right)
\chi'_\eta \nonumber \\
&& -\half\theta'_\eta\left( \kappa_5(\eta\cdot\zeta)(\eta'\cdot\zeta) +
\frac{\kappa_5-4}{2}\right) \chi_\eta + \half\theta_\eta \theta'_\eta
\left( \kappa_5(\eta\cdot\zeta)(\eta'\cdot\zeta) +
\frac{\kappa_5}{2}\right)B.
\eea
We demand that $\Ng \psi = \ell \psi$ for some eigenvalue
$\ell$, therefore we can write
\[
\kappa_5\left( \half(\eta\cdot\zeta)(\eta'\cdot\zeta) + \frac{1}{4}
\right) A = \ell A,
\]
\ie
\beq
\kappa_5(\eta\cdot\zeta)(\eta'\cdot\zeta) A = \left(\frac{4\ell
-\kappa_5}{2}\right) A.
\eeq
Similarly
\bea
\kappa_5(\eta\cdot\zeta)(\eta'\cdot\zeta) \chi'_\eta &=&
\left(\frac{4\ell -\kappa_5 -4}{2} \right) \chi'_\eta,
\label{eq:chi'eig}\\
\kappa_5(\eta\cdot\zeta)(\eta'\cdot\zeta)\chi_\eta &=&
\left(\frac{4\ell -\kappa_5 +4}{2} \right) \chi_\eta,
\label{eq:chieig}\\
\kappa_5(\eta\cdot\zeta)(\eta'\cdot\zeta)B &=& \left(\frac{4\ell
-\kappa_5}{2}\right) B. \label{eq:Beig}
\eea
In Appendix A the diagonalisation of
$(\eta\cdot\zeta)(\eta'\cdot\zeta)$ is carried out explicitly in the
occupation number basis $\{ |m,n \rangle \}$. Below we assume that
suitable eigenstates can be found, and explore the consequences for
the cohomology of the BRST operator at generic ghost number $\ell$.

\subsection{Action of BRST operator}

The BRST charge is defined above as $\Omega = \eta^\alpha L_{\alpha-}$ and
from Section \ref{sec:realis} we can write
\beq
L_{\alpha-} = \theta_\alpha P_- - X_- \frac{\partial}{\partial
\theta^\alpha} - \zeta_\alpha \zeta_-.
\eeq
We have previously \eqn{eq:xi-} defined $\zeta_-$ and can
write it as $\zeta_- =\frac{1}{2P_+} \left( D_5 (-1)^z +
\kappa_5\zeta^\beta P_\beta \right)$, where $D_5 =  \gamma_5(\gamma
\cdot P - \M)$ is the Dirac operator multiplied by $\gamma_5$.
Consequently
\bea
\eta^\alpha\zeta_\alpha\zeta_- &=& \frac{\eta^\alpha
\zeta_\alpha}{2P_+} (D_5 (-1)^z + \kappa_5\zeta^\beta P_\beta),
\nonumber \\
&=&\frac{\eta^\alpha\zeta_\alpha (-1)^z D_5}{2P_+} +
\kappa_5\frac{(\eta\cdot\zeta)\left(\zeta^\beta
\frac{\partial}{\partial \theta^\beta}\right)}{2P_+}.
\label{eq:zetaa-}
\eea
The second part of equation \eqn{eq:zetaa-} can be further expanded
as follows
\beq
\zeta^\beta \frac{\partial}{\partial \theta^\beta} = \zeta_\gamma
\frac{\partial}{\partial \theta^\beta} \eps^{\beta\gamma} =
\zeta_\gamma \frac{\partial}{\partial \theta^\beta}(-\eta^\beta
\eta'^\gamma + \eta^\gamma \eta'^\beta),
\eeq
which uses the identity $\eps^{\beta\alpha} = (-\eta^\beta
\eta'^\alpha + \eta^\alpha \eta'^\beta)$. Therefore
\beq
\frac{(\eta\cdot\zeta)\left(\zeta^\beta \frac{\partial}{\partial
\theta^\beta} \right)}{2P_+} =  \frac{1}{2P_+}
(\eta\cdot\zeta)(\eta'\cdot\zeta) \frac{\partial}{\partial
\theta'_\eta} + \frac{1}{2P_+} (\eta\cdot\zeta)^2
\frac{\partial}{\partial \theta_\eta},
\eeq
and so we can write
\beq
\eta^\alpha\zeta_\alpha\zeta_- = \frac{\eta^\alpha\zeta_\alpha (-1)^z
D_5}{2P_+} + \frac{\kappa_5}{2P_+} (\eta\cdot\zeta)(\eta'\cdot\zeta)
\frac{\partial}{\partial \theta'_\eta} + \frac{\kappa_5}{2P_+}
(\eta\cdot\zeta)^2 \frac{\partial}{\partial \theta_\eta}.
\eeq

In a similar fashion we can expand the fermionic part of
$L_{\alpha-}$ as follows
\[
\eta^\alpha L_{\alpha-}^L = \eta^\alpha \theta_\alpha P_- -
\eta^\alpha X_- \frac{\partial}{\partial \theta^\alpha} = \theta_\eta
P_- + \frac{\partial}{\partial \theta'_\eta} X_-,
\]
and
\[
P_- = \frac{-1}{2P_+} \left( (P^2 - \M^2) + Q^\alpha Q_\alpha\right),
\]
but
\bea
\eps^{\beta\alpha}Q_\alpha Q_\beta = Q^\alpha Q_\alpha &=&
(-\eta^\alpha\eta'^\beta + \eta^\beta\eta'^\alpha)
\frac{\partial}{\partial\theta^\alpha}\frac{\partial}{\partial\theta^\beta},
\nonumber \\
&=& -2\frac{\partial}{\partial \theta_\eta}\frac{\partial}{\partial
\theta'_\eta}, \nonumber
\eea
therefore
\beq
P_- = \frac{-1}{2P_+} \left((P^2 - \M^2) + 2\frac{\partial}{\partial
\theta_\eta}\frac{\partial}{\partial \theta'_\eta} \right).
\label{eq:KGeqn2}
\eeq
The BRST operator can thus be written
\bea
\Omega &=& \eta^\alpha L_{\alpha-} = -\theta_\eta \frac{P^2 -
\M^2}{2P_+} - \frac{\theta_\eta}{P_+}\frac{\partial}{\partial
\theta_\eta}\frac{\partial}{\partial \theta'_\eta} +
\frac{\partial}{\partial \theta'_\eta} X_-, \nonumber\\
& & -\frac{\eta^\alpha \zeta_\alpha (-1)^z D_5}{2P_+} -
\kappa_5\frac{(\eta\cdot\zeta)(\eta'\cdot\zeta)}{2P_+}\frac{\partial}{\partial\theta'_\eta}
- \kappa_5\frac{(\eta\cdot\zeta)^2}{2P_+}\frac{\partial}{\partial \theta_\eta}.
\eea
By writing $\psi$ as a series expansion to second order (equation
\eqn{eq:psispin}), we can determine the effect of $\Omega$ on $\psi$.
For simplicity we shall write the effect of each term of $\Omega$ on
$\psi$ separately.
\newline
$1^{st}$ Term:
\[
-\theta_\eta\frac{P^2 - \M^2}{2P_+} \psi = -\theta_\eta\frac{P^2 -
\M^2}{2P_+} A + \theta_\eta \theta'_\eta \frac{P^2 - \M^2}{2P_+}
\chi_\eta.
\]
$2^{nd}$ Term:
\[
-\frac{\theta_\eta}{P_+}\frac{\partial}{\partial
\theta_\eta}\frac{\partial}{\partial \theta'_\eta} \psi = \frac{2
\theta_\eta B}{2P_+}.
\]
$3^{rd}$ Term:
\[
\frac{\partial}{\partial \theta'_\eta} X_- \psi = -X_- \chi_\eta -
\theta_\eta X_- B.
\]
$4^{th}$ Term:
\bea
-\frac{\eta^\alpha \zeta_\alpha (-1)^z D_5}{2P_+} \psi &=& -(\eta
\cdot \zeta) \frac{D_5}{2P_+} A + \theta_\eta (\eta\cdot\zeta)
\frac{D_5}{2P_+}\chi'_\eta \nonumber \\
&& \;\;\;- \theta'_\eta (\eta\cdot\zeta)\frac{D_5}{2P_+}\chi_\eta -
\theta_\eta\theta'_\eta (\eta\cdot\zeta) \frac{D_5}{2P_+}B. \nonumber
\eea
$5^{th}$ Term:
\[
-\kappa_5\frac{(\eta\cdot\zeta)(\eta'\cdot\zeta)}{2P_+}\frac{\partial}{\partial\theta'_\eta}
\psi = \frac{\kappa_5}{2P_+} (\eta\cdot\zeta)(\eta'\cdot\zeta)
\chi_\eta + \kappa_5\frac{\theta_\eta}{2P_+}
(\eta\cdot\zeta)(\eta'\cdot\zeta) B.
\]
$6^{th}$ Term:
\[
-\kappa_5\frac{(\eta\cdot\zeta)^2}{2P_+}\frac{\partial}{\partial
\theta_\eta} \psi = -\frac{\kappa_5}{2P_+} (\eta\cdot\zeta)^2
\chi'_\eta - \kappa_5 \theta'_\eta - \frac{\theta'_\eta}{2P_+}
(\eta\cdot\zeta)^2 B
\]
Grouping $\Omega\psi$ with respect to coefficients of $\theta_\eta,
\theta'_\eta$ and $\theta_\eta\theta'_\eta$ we can write
\beq
\Omega\psi = C + C_{\theta_\eta} \theta_\eta +
C_{\theta'_\eta}\theta'_\eta +
C_{\theta_\eta\theta'_\eta}\theta_\eta\theta'_\eta
\eeq
where we have
\beq
C = -X_- \chi_\eta - \eta\cdot\zeta \frac{D_5}{2P_+}A +
\frac{\kappa_5}{2P_+}(\eta\cdot\zeta)(\eta'\cdot\zeta)\chi_\eta -
\frac{\kappa_5}{2P_+}(\eta\cdot\zeta)^2 \chi'_\eta,
\label{eq:c}
\eeq
\vspace{-3mm}
\beq
C_{\theta_\eta} = -\frac{P^2 - \M^2}{2P_+} A + \frac{B}{P_+} - X_- B
+ \eta\cdot\zeta \frac{D_5}{2P_+}\chi'_{\eta} + \frac{\kappa_5}{2P_+}
(\eta\cdot\zeta)(\eta'\cdot\zeta)B,
\label{eq:th}
\eeq
\vspace{-5mm}
\bea
C_{\theta'_\eta} &=& -\eta\cdot\zeta \frac{D_5}{2P_+} \chi_\eta -
\kappa_5(\eta\cdot\zeta)^2\frac{B}{2P_+}, \label{eq:th'}\\
C_{\theta_\eta\theta'_\eta} &=& \frac{P^2 - \M^2}{2P_+} \chi_\eta -
\eta\cdot\zeta \frac{D_5}{2P_+} B. \label{eq:thth'}
\eea
Notice the apparent similarity between equations \eqn{eq:th'} and
\eqn{eq:thth'}, these can in fact be shown to be a linear
transformation of each other. Firstly, note that 
\beq
D^2_5 = \gamma_5(\gamma\cdot P - \M)\gamma_5(\gamma\cdot P - \M = - \kappa_5
(P^2 - \M^2),
\nonumber
\eeq thus we can write equation \eqn{eq:thth'}
\bea
C_{\theta_\eta \theta'_\eta}&=& -\frac{D_5^2}{2\kappa_5 P_+} \chi_\eta -
\frac{(\eta\cdot\zeta)D_5}{2P_+} B, \nonumber \\
&=& -\frac{D_5}{\kappa_5(\eta\cdot\zeta)} \left(
\frac{(\eta\cdot\zeta) D_5}{2P_+}\chi_\eta +
\kappa_5\frac{(\eta\cdot\zeta)^2}{2P_+}B \right).
\eea
Thus it can be seen that equations \eqn{eq:th'} and \eqn{eq:thth'}
differ only by a factor of $-D_5/(\kappa_5(\eta\cdot\zeta))$.
Note that if
\[
\frac{(\eta\cdot\zeta) D_5}{2P_+}\chi_\eta +
\kappa_5\frac{(\eta\cdot\zeta)^2}{2P_+}B= 0,
\]
then both equations \eqn{eq:th'} and \eqn{eq:thth'} will be zero.
It is interesting to note that a similar situation exists in
equations \eqn{eq:c} and \eqn{eq:th}. The common component of these
two equations is
\[
\frac{D_5}{2P_+}A + \kappa_5\frac{(\eta\cdot\zeta)}{2P_+}\chi'_\eta.
\]
Taking these similarities between the two pairs of equations into
account  we can redefine the expansion of $\psi$ as follows: rescale
$\chi_\eta$ and $A$ by
\bea
\chi_\eta &\equiv& \tilde{\chi}_\eta + (\eta\cdot\zeta)\frac{D_5}{P^2
- \M^2} B, \label{eq:chitilde} \\
A &\equiv& \tilde{A} + (\eta\cdot\zeta)\frac{D_5}{P^2 -\M^2}
\chi'_\eta \label{eq:Atilde},
\eea
and so $\psi$ becomes
\beq
\psi = \left(\tilde{A} + (\eta\cdot\zeta)\frac{D_5}{P^2 -\M^2}
\chi'_\eta\right) + \theta_\eta \chi'_\eta - \theta'_\eta
\left(\tilde{\chi}_\eta +
(\eta\cdot\zeta)\frac{D_5}{P^2-\M^2}B\right) + \theta_\eta
\theta'_\eta B.
\eeq
Using this redefinition equation \eqn{eq:th'} becomes
\bea
C_{\theta'_\eta} &=& -\eta\cdot\zeta \frac{D_5}{2P_+}
\tilde{\chi}_\eta - \frac{(\eta\cdot\zeta)^2}{2P_+} \frac{D_5^2}{P^2
- \M^2} B - \kappa_5\frac{(\eta\cdot\zeta)^2}{2P_+} B, \nonumber \\
 &=& -\eta\cdot\zeta \frac{D_5}{2P_+} \tilde{\chi}_\eta
+\frac{(\eta\cdot\zeta)^2}{2P_+} \frac{\kappa_5(P^2 - \M^2)}{P^2 -
\M^2} B - \kappa_5\frac{(\eta\cdot\zeta)^2}{2P_+} B, \nonumber \\
&=& -\eta\cdot\zeta \frac{D_5}{2P_+} \tilde{\chi}_\eta.
\label{eq:ctheta'}
\eea
By enforcing $\Omega \psi = 0$ we get $C_{\theta'_\eta} = 0$, which by \eqn{eq:ctheta'} gives the Dirac equation
\beq
-\eta\cdot\zeta \frac{D_5}{2P_+} \tilde{\chi}_\eta = 0.
\label{eq:diracchitilde}
\eeq
Similarly we can rewrite equation \eqn{eq:thth'} as
\beq
C_{\theta_\eta\theta'_\eta} = \frac{P^2 - \M^2}{2P_+}
\tilde{\chi}_\eta = \frac{1}{\kappa_5}D_5(D_5 \tilde{\chi}_\eta),
\eeq
which, by enforcing $\Omega \psi = 0$ leads to the Klein-Gordon equation
\beq
D_5^2 \tilde{\chi}_\eta = 0.
\eeq
Under the rescaling of equation \eqn{eq:Atilde} equation \eqn{eq:c}
becomes
\bea
C &=& \left(-X_- +
\frac{\kappa_5}{2P_+}(\eta\cdot\zeta)(\eta'\cdot\zeta)\right)\chi_\eta
- (\eta\cdot\zeta) \frac{D_5}{2P_+} \tilde{A} +
\kappa_5\frac{(\eta\cdot\zeta)^2}{2P_+}\chi'_\eta  \nonumber \\
& &-\kappa_5\frac{(\eta\cdot\zeta)^2}{2P_+}\chi'_\eta,
\label{eq:cagain}
\eea
and by substituting equation \eqn{eq:chieig} into \ref{eq:cagain} we get
\beq
C=\left(-X_- + \frac{4\ell -\kappa_5 + 4}{4P_+} \right)\chi_\eta -
(\eta\cdot\zeta) \frac{D_5}{2P_+} \tilde{A} = 0.
\label{eq:1}
\eeq
Similarly, using equations \eqn{eq:Atilde} and \eqn{eq:Beig}, \eqn{eq:th} can now be written
\beq
C_{\theta_\eta} = \frac{D_5^2}{2\kappa_5 P_+} \tilde{A} - \left(X_- -
\frac{4\ell - \kappa_5 + 4}{4P_+} \right) B = 0,
\label{eq:2}
\eeq
where once again we have enforced the condition for physical states.

Defining the symbol
\beq
\tilde{X}_- = X_- - \frac{4\ell - \kappa_5 + 4}{4P_+},
\eeq
and multiplying equation \eqn{eq:2} by
$(\eta\cdot\zeta)\frac{D_5}{P^2 -\M^2}$ and then subtracting it from
equation \eqn{eq:1} we get
\beq
-\tilde{X}_- \left(\chi_\eta - (\eta\cdot\zeta)\frac{D_5}{P^2 - \M
^2} B \right) = 0.
\label{eq:xtilde}
\eeq
Substituting equation \eqn{eq:chitilde} into \eqn{eq:xtilde} gives the third equation of motion
\beq
-\tilde{X}_- \tilde{\chi}_\eta = 0.
\label{eq:third}
\eeq

Finally, we identify the physical states at generic ghost number $\ell$, arising from the cohomology of $\Omega$,  as the spinors $\tilde{\chi}$, with the following properties.
From \eqn{eq:diracchitilde}, the $\tilde{\chi}$ obey the usual massive
Dirac equation. From \eqn{eq:KGeqn},\eqn{eq:KGeqn2}, the
$P_{-}$ constraint dictates the dependence of superfield components on
light cone time $\tau = x^{-}$, as $P_{-} = \partial/\partial x^{-}$.
Finally, interpreting \eqn{eq:xtilde}, \eqn{eq:third} in the
$p_{+}$-representation (the Fourier transform of the
$x^{+}$-representation, \ie $X_{-}\equiv X^{+} = -\partial/\partial p_{+}$),
the $\tilde{\chi}$ are homogeneous functions of $p_{+}$ of degree
$-(+1 -\frac 14 \kappa_{5}+\ell)$. Thus the $\tilde{\chi}$ are
essentially only functions over $(d-1)+1$-dimensional Minkowski space. For example, in the case $\kappa_5 = 1, \ell = -3/4$ (which implies $\Lambda = 0$), and using \eqn{eq:lambdaseries}, we find that the physical states have the following explicit form\footnote{$-2\ell = 3/2$ is the correct conformal dimension for a spinor field (see \cite{macksalam})}
\beq
|\tilde{\chi}_\eta \rangle = \tilde{\chi}_\eta (x^\mu) \left[|0,0\rangle - |1,1\rangle + |2,2\rangle + \ldots + (-1)^m | m,m\rangle + \ldots\right],
\eeq
in terms of the number states of the bosonic ghost sector (see Appendix A), where $\tilde{\chi}_\eta(x^\mu)$ are ordinary functions of $x^\mu$.

\section{BRST-BFV quantisation of the spinning particle and
$iosp(d,2/2)$ structure}
\label{brstbfv}

As is well known \cite{govbk,henneaux}, the BFV canonical quantisation
of constrained Hamiltonian systems \cite{BFV1,BFV2,BFV3} uses an extended phase
space description in which, to each first class constraint $\phi_a$,
a pair of conjugate `ghost' variables (of Grassmann parity opposite
to that of the constraint) is introduced. Here we follow this
procedure for the spinning relativistic particle. Although our
notation is adapted to the massive case, $\M >0$, as would follow
from the second order action corresponding to extremisation of the
proper length of the particle world line, an analysis of the {\it
fundamental} Hamiltonian description of the first order action\cite
{govbk} leads to an equivalent picture (with an additional mass
parameter $\mu \ne 0$ supplanting $m$ in appropriate equations, and
permitting $m\rightarrow 0$ as a smooth limit).

In either case, for the scalar or spinning particle the primary first class constraint is the
mass-shell condition $\phi_1 = (P^2 - \M^2)$, where $P^2 =
P^{\mu}\eta_{\mu\nu}P^{\nu}$. Including the Lagrange multiplier
$\lambda$ as an additional dynamical variable leads to a secondary
constraint, reflecting conservation of its conjugate momentum
$\pi_\lambda$.
The quantum formulation should be
consistent with the equations of motion and gauge fixing at the
classical level, as such two restrictions are necessary so as to
arrive at the particle quantisation corresponding with the
superalgebraic prescription of Section \ref{iosp}. Firstly, we choose
below to work in the class \cite{monaghan,teitelboim1,teitelboim2,teitelbk}
$\dot{\lambda}=0$; moreover, we take gauge fixing only to be with
respect to gauge transformations in one of the {\it connected}
components of the group, \ie either the identity class, or the
orientation reversing class characterised by $\tau' = \tau_i + \tau_f
- \tau$. Thus $\lambda$ will be quantised on the half line (say
$\IR^+$), and the system is not modular invariant until the two
distinctly oriented sectors (particle and anti-particle) are combined
\cite{govbk}. Secondly, we take $\phi^{(2)}_1 = \lambda \pi_\lambda$ as
the other secondary first class constraint (rather than $\phi^{(2)}_1 =
\pi_\lambda$ used in the standard construction)
\footnote{We thank J Govaerts for clarifying possible difficulties with
regularity and independence of constraints}. Finally, the spinning particle system also entails a second, Grassman odd first class constraint $\phi_2 = p_\mu \zeta^\mu + \M \epsilon \gamma_5 \; (\epsilon = \pm 1)$, together  with its associated first class constraint $\phi_2^{(2)} = \pi_2^{(2)}$, the conjugate momentum of the corresponding Lagrange multiplier $\lambda_2$ (which is also Grassmann odd).

\subsection{BFV extended state space and wavefunctions}

The BFV extended phase space \cite{govbk} for the BRST quantisation
of the spinning relativistic particle is therefore taken to comprise the
following canonical variables:
\beq
x^\mu(\tau), p_\mu(\tau), \; \zeta^\mu, \zeta_5, \; \lambda(\tau),
\pi_\lambda(\tau), \; \lambda_2(\tau), \pi_2(\tau), \; \eta^{a(i)},
\rho_{a(i)}, \; \; a,i = 1,2.
\label{eq:variables}
\eeq
$x^\mu(\tau), p_\mu(\tau)$ are Grassmann even whilst $\zeta^\mu,
\zeta_5$ are Grassmann odd variables, $\lambda$ is the Grassmann even
Lagrange multiplier corresponding to the even first class constraint
$\phi_1$, $\pi_\lambda$ is the momentum conjugate to $\lambda$ (which
forms the constraint $\phi^{(2)}_1$), $\lambda_2$ is the odd
Lagrange multipler corresponding to the  Grassmann odd first class
constraint $\phi_2$, and
$\phi_2^{(2)} = \pi^{(2)}_2$ is its conjugate momentum. $\eta^{1(1)}, \rho_{1(1)}$ and $\eta^{1(2)},
\rho_{1(2)}$ are the Grassmann odd conjugate pairs of ghosts
corresponding to the constraints $\phi_1$ and $\phi^{(2)}_1$
respectively, while $\eta^{2(1)}, \rho_{2(1)}$ and $\eta^{2(2)},
\rho_{2(2)}$ are the Grassmann even conjugate pairs of ghosts
corresponding to the constraints $\phi_2$ and $\phi^{(2)}_2$
respectively. We proceed directly to the quantised version by
introducing the Schr\"{o}dinger representation. We
introduce  operators $X^\mu, P_\nu$ corresponding to the coordinates
$x^\mu, p_\nu$, acting on suitable sets of wavefunctions over
$x^{\mu}$, and on the half
line $\lambda >0$. The Hermitian ghosts $\eta^{a(i)}, \rho_{b(j)}$ (a pair
of bc
systems) are represented as usual either on a 4-dimensional indefinite inner
product space $|\sigma\sigma'\rangle, \; \sigma,\sigma' = \pm$, or here, in
order to match with Section \ref{iosp}, in terms of suitable Grassmann variables acting on superfields. The non zero commutation relations amongst \eqn{eq:variables} read
(repeated in full for clarity):
\beq
\begin{array}{c}
\begin{array}{ccc}
[X_\mu,P_\nu] = -\eta_{\mu\nu}, & \{\zeta_\mu,\zeta_\nu \} =
-\frac{2}{\kappa_5}\eta_{\mu\nu}, & \{\gamma_5,\gamma_5 \} =
2\kappa_5,
\end{array}
\\
\begin{array}{cc}
[\lambda,\pi_\lambda] = 1, & \{\lambda_2, \pi_2\} = -1,
\end{array}
\\
\begin{array}{ccc}
\{\eta^{1(i)},\rho_{1(j)} \} = -\delta^i_j, &
[\eta^{2(i)},\rho_{2(j)} ] = \delta^i_j & i,j = 1,2,
\end{array}
\end{array}
\label{eq:algmat}
\eeq
from which the algebra of constraints follows:
\beq
\begin{array}{cc}
\{\phi_2,\phi_2\} = - 2 \kappa_5\phi_1 & \{\phi_1,\phi_2 \} = \{\phi_1,\phi_1\}
= 0.
\end{array}
\eeq
The Hermiticity conditions imposed on the above operators read
\beq
\begin{array}{c}
\begin{array}{cccc}
X_\mu ^\dag = X_\mu , & P_\mu ^\dag =P_\mu ,&
\zeta_\nu^\dagger=\zeta_\nu ,& \mu,\nu =0,...d-1,  \nonumber \\
\zeta_0^\dagger =-\zeta_0,& \gamma_5^{\dagger }=\gamma_5, & & \\
\lambda^\dag = \lambda, & \pi_\lambda^\dag=\pi_\lambda, &
\lambda_2^\dagger = -\lambda_2,& \pi_2^\dagger=\pi _2,
\end{array}
\\
\begin{array}{ccc}
(\eta^{a(i)})^\dagger =\eta^{a(i)},&
(\rho_{a(i)})^\dag=-(-1)^{a+1}(\rho_{a(i)}),& a=1,2.
\end{array}
\end{array}
\eeq
The ghost number operator $\Ng$ is defined by
\beq
\Ng = \half \sum_{a,i =1}^2 \left( \eta^{a(i)}\rho_{a(i)} -
(-1)^{(a-1)} \rho_{a(i)}\eta^{a(i)}\right).
\eeq
The canonical BRST operator\footnote{The criteria for the construction and nilpotency of the corresponding anti-BRST operator $\bar{\Omega}$ have been given in \cite{hwang}} is given by
\beq
\Omega = \eta^{1(1)} \phi_1 + \eta^{1(2)} \phi^{(2)}_1 + \eta^{2(1)}
\phi_2 + \eta^{2(2)} \phi^{(2)}_2 + \half\left(\eta^{2(1)}\right)^2
\rho_{1(1)}.
\eeq
The gauge fixing operator \cite{monaghan} ${\cal F}$ which will lead to
the appropriate effective Hamiltonian is given by:
\beq
{\cal F} = - \half \lambda \rho_{1(1)},
\label{eq:gaugeferm}
\eeq
and thus the Hamiltonian can be written
\beq
H = -\gbl{\cal F},\Omega\gbr = -\half\lambda\left(\eta^{1(2)}\rho_{1(1)} +
\phi_1 \right),
\label{eq:hamil}
\eeq
which is of course BRST invariant.

Consider the following canonical transformations on the classical
dynamical variables of the extended phase space \cite{casalbuoni}
\beq
\begin{array}{rcl}
\eta'^{a(i)} &=& \lambda \eta^{a(i)}, \\
\rho'_{a(i)} &=& \frac{1}{\lambda} \rho_{a(i)}, \\
\end{array}
a(i) = 1(1), 1(2), 2(1),
\label{eq:canon1}
\eeq
\beq
\lambda \pi'_\lambda = \lambda \pi_\lambda + (\eta^{1(1)}\rho_{1(1)} +
\rho_{1(2)}\eta^{1(2)} - \eta^{2(1)}\rho_{2(1)}),
\label{eq:canon2}
\eeq
with the remainder invariant. At the same time we relabel the
co-ordinates $p_+ = \lambda^{-1}$ and $x_- = \lambda \pi_\lambda
\lambda$. At the quantum level the corresponding BRST operator
$(\Omega' = \eta'^{1(1)} \phi_1 + \eta'^{1(2)} \phi'^{(2)}_1 +
\eta'^{2(1)} \phi_2 + \eta^{2(2)} \phi^{(2)}_2 +
\half\left(\eta'^{2(1)}\right)^2 \rho'_{1(1)})$ can be written as
\bea
\Omega' & = & \lambda \eta^{1(1)} \phi_1 + \eta^{1(2)} :\lambda
\phi_1^{(2)} : + \lambda \eta^{2(1)}\phi_2 + \eta^{2(2)} \phi_2^{(2)}
\nonumber\\
& & - \lambda \eta^{1(2)}\eta^{1(1)} \rho_{1(1)} - \lambda
\eta^{1(2)} \eta^{2(1)}\rho_{2(1)} + \half \lambda \left( \eta^{2(1)}
\right)^2 \rho_{1(1)},
\eea
where the symmetric ordering
\[
:\lambda \phi_1^{(2)} : = \half (\lambda \phi_1^{(2)} + \phi_1^{(2)}
\lambda ) = \lambda \phi_1^{(2)} - \half \lambda,
\]
has been introduced.

It is also convenient to define \cite{casalbuoni} the operators
$\theta_\alpha, Q_\alpha, \zeta'_\alpha$ and $\tilde{\zeta}'_\alpha \;
(\alpha = 1,2)$ by
\beq
\begin{array}{rclrcl}
Q_{1,2} &=& \frac{1}{2\sqrt{2}} \left(2\eta^{1(2)} \pm \rho_{1(1)}
\right), &
     \theta_{1,2} & = & \frac{1}{\sqrt{2}} \left( \pm \rho_{1(2)} -
2\eta^{1(1)} \right), \\
\zeta'_{1,2} &=& \frac{1}{\sqrt{2}}\left( \eta^{2(1)} \pm \rho_{2(1)} \right), &
   \tilde{\zeta}'_{1,2} & = &  \frac{1}{\sqrt{2}}\left( \pm \eta^{2(2)} - \rho_{2(2)}
\right),
\end{array}
\eeq
which obey the relations
\bea
\{ Q_\alpha,X_\beta \} = \eps_{\alpha\beta} &\mbox{and}& 
[\zeta'_\alpha,\zeta'_\beta ] = - \eps_{\alpha\beta}.
\eea
In terms of these variables we attain the following simple forms for
the BRST, gauge fixing and Hamiltonian operators.
\bea
\Omega ' &=& \frac{1}{\sqrt{2}} \left( :\lambda \phi_1^{(2)} : (Q_1 +
Q_2) + (\theta_1 + \theta_2) H +  (\zeta'_1 + \zeta'_2) \lambda
(\phi_2 + Q^\alpha \zeta'_\alpha) + (\tilde{\zeta}'_1
- \tilde{\zeta}'_2 )\phi_2^{(2)} \right), \nonumber \\
{\cal F}' &=& - \half \rho_{1(1)} = -\frac{1}{\sqrt{2}} (Q_1 - Q_2), \label{eq:allthree}\\
H' &=& - \gbl {\cal F}',\Omega'\gbr = - \frac{\lambda}{2} \left( P^\mu
P_\mu + Q^\alpha Q_\alpha - \M^2 \right) \equiv H. \nonumber
\eea
Note that the $\zeta'_\alpha$ defined here and the $\zeta_\alpha$ defined in section \ref{sec:realis} differ by a factor $\sqrt{2}$, \ie $\zeta'_\alpha = \frac{1}{\sqrt{2}} \zeta_\alpha$.

\subsection{$\beta-$limiting procedure for the BRST operator}
\label{betalimit}

It is now necessary to reconcile the development of Sections \ref{iosp} and \ref{physstates},
in which the identical raw material for construction of the BRST
operator, gauge fixing function and hence physical states, appears
{\it purely algebraically} (compare equations \eqn{eq:rawmat1},\eqn{eq:rawmat2},\eqn{eq:rawmat3} with \eqn{eq:algmat})
except for the absence of the $\eta^{2(2)},\rho_{2(2)}$ even ghosts and
thus the
$\tilde{\zeta}_1, \tilde{\zeta}_2$ oscillators. In \cite{casalbuoni},
a somewhat heuristic argument was
provided to justify the restriction to the vacuum of the
latter oscillators. Here instead we shall use what is
known as the $\beta-$limiting procedure \cite{govbk} applied
throughout on the $a=2$ label of the BFV phase space variables (if we
also apply it to the $a=1$ label we recover the Fadeev-Popov
reduced phase space quantisation scheme). The exposition will closely
follow that of
\cite{govbk}.

Consider instead of \eqn{eq:hamil} the gauge fixing fermion
\beq
{\cal F} = -\half \lambda \rho_{1(1)} + \frac{1}{\beta}(\lambda_2 -
\lambda_2^0)\rho_{2(2)} + \lambda_2\rho_{2(1)},
\eeq
where $\beta$ is arbitrary, real and
Grassmann even and $\lambda_2^0$ is some given function of time with
the same properties as $\lambda_2$. The Hamiltonian is thus given by
\bea
\Hf = \gbl {\cal F},\Omega \gbr &=& - \half \lambda (\phi_1 +
\eta^{1(2)}\rho_{1(1)}) + \frac{1}{\beta}(\lambda_2 -\lambda_2^0)
\phi_2^{(2)} + \frac{1}{\beta} \eta^{2(2)}\rho_{2(2)} \nonumber\\
&&+ \eta^{2(2)}\rho_{2(1)} + \lambda_2 \eta^{2(1)}\rho_{1(1)} + 
\lambda_2 \phi_2.
\eea
The equations of motion for the BFV phase space variables can be
easily obtained for the above $H$ by implementing as usual $\dot{A} =
\gbl A,\Hf \gbr$. We now change to new variables $\tilde{\pi}_2,
\tilde{\rho}_{2(2)}$ such that $\pi_2 = \beta \tilde{\pi}_2$ and
$\rho_{2(2)} = \beta \tilde{\rho}_{2(2)}$ and subsitute these into
$\Hf, \Omega$ and $\Ng$, the equations of motion and the action
related to $\Hf$. Having done that we take the limit $\beta \ra 0$,
in particular the BRST and ghost number operators then become
\bea
\Omega & = & \eta^{1(1)} \phi_1 + \eta^{1(2)} \phi_1^{(2)} +
\eta^{2(1)} \phi_2 + \half (\eta^{2(1)})^2
\rho_{1(1)},\label{eq:betabrst} \\
\Ng & = & \sum_{i=1}^{2} \eta^{1(i)}\rho_{1(i)} +
\eta^{2(1)}\rho_{2(2)} - \half\,
\label{eq:betaghost}
\eea
while the equations of motion for $\lambda_2, \pi_2, \eta^{2(2)}$ and
$\rho_{2(2)}$ (which are the ones that are affected by the
$\beta-$limiting procedure) now become:
\beq
\begin{array}{cccc}
(\lambda_2 - \lambda_2^0) = 0, & \tilde{\pi}_2 = - \phi_2 -
\eta^{2(1)} \rho_{1(1)}, & \eta^{2(2)} = 0, & \rho_{2(2)} =
-\tilde{\rho}_{2(2)}.
\end{array}
\eeq
Solving these equations and taking $\lambda_2^0 = 0$, we find that
equations \eqn{eq:betabrst} and \eqn{eq:betaghost} remain as they are
whilst $\Hf = H$. Thus we have succeeded in `squeezing out' the
$2(2)$ pair of even ghosts together with the odd Lagrange multiplier
$\lambda_2$. Moreover the Hamiltonian in equation \eqn{eq:hamil},
obtained from the admissible gauge fixing fermion given in equation
\eqn{eq:gaugeferm}, is recovered.

Finally, and most importantly, the canonical transformation in
equations \eqn{eq:canon1}, \eqn{eq:canon2} is not affected by this
procedure, as can easily be observed. Thus whether we apply the
canonical transformations before the $\beta-$limiting procedure or
after does not matter. Consequently via equations \eqn{eq:canon1},
\eqn{eq:canon2} equation \eqn{eq:betabrst} becomes
\beq
\Omega' =  \frac{1}{\sqrt{2}} \left( :\lambda \phi_1^{(2)} : (Q_1 +
Q_2) + (\theta_1 + \theta_2) H +  (\zeta'_1 + \zeta'_2) \lambda
(\phi_2 + \ Q^\alpha \zeta_\alpha) \right),
\label{eq:BRST}
\eeq

The forms \eqn{eq:hamil} and \eqn{eq:BRST} can now be shown to be identical to the previously given algebraically defined expressions for these quantities (\eqn{eq:KGeqn}, \eqn{eq:BRSTandPsi}. The raw material \eqn{eq:rawmat1},\eqn{eq:rawmat2},\eqn{eq:rawmat3} also appears in this construction, as can be easily observed by \eqn{eq:algmat}, and by identifying $P_+ = \lambda^{-1}, X_- = :\lambda \phi_1^{(2)} :, \zeta_- = \phi_2 + Q^\alpha\zeta'_\alpha$ and the BRST operator $\Omega' = \eta^\alpha L_{\alpha-}$. Moreover, the realisation of $iosp(d,2/2)$ can be done as in \eqn{eq:jalpha-}, \eqn{eq:js}. In particular, the evaluation of the BRST cohomology performed in Section \ref{physstates} above, gives precisely the correct identification of physical state wavefunctions for the spinning particle model of this section, provided that we represent $\zeta'_\alpha$ by $(-1)^z \zeta'_\alpha$ to account for the correct action on the superfield and the correct commutation relations of $iosp(d,2/2)$. Also, the constant $\kappa_5$ appearing in \eqn{eq:KGeqn} can also be introduced in the third equation of \eqn{eq:allthree} to account for $\gamma_5 = \pm 1$, which will eventually appear in the factorisation of $P_-$ leading to the Dirac equation.

\section{Conclusions}
\label{conc}

The present work, via the positive results
claimed here for the test case of the spinning particle,
provides confirmation of our programme of
establishing the roots of covariant quantisation of relativistic
particle systems, in the BRST complex associated with representations of classes
of extended spacetime supersymmetries. Similar
examples under study are the `$D(2,1; \alpha)$' particle in $1+1$
dimensions\cite{d21agrp22}, the higher spin-$s$ case and the
relation to Bargmann-Wigner equations, as well as considerations of
how the method can be extended to, say, superstring or superparticle
cases, for which a covariant approach has so far been
problematical\cite{kappasusy}. The general approach\cite{wigsym,clausthal} is
a classification of `quantisation superalgebras' in diverse
dimensions, whose representation theory will implement the covariant
quantisation, in the spirit of the above example, of the appropriate
classical phase space models of systems with gauge symmetries.

Conformal (super)symmetry has long been of interest as a probable
higher symmetry underlying particle interactions, no more so than
in the light of recent interpretations of compactifications of higher dimensional supergravities \cite{maldacena,witten}. The present application is of particular
interest in that the traditional descent from $d+2$ to $d$
dimensions - via a projective conformal space \cite{macksalam} - is here
implemented not on the cone (massless irreps), but for the massive
(super)hyperboloid. The present work can also be seen as an
elaboration of the method of `conformalisation' \cite{siegel}, and as a version of `two time' physics \cite{bars1,bars2,bars3}.
Beyond the Dirac equation and higher spin generalisations, it will
also be possible to investigate the algebraic BRST-BFV complex
associated with indecomposable representations\cite{bracken1,bracken2} (for example
where the
vector-scalar (super) special
conformal generators are represented as nilpotent matrices).

Finally, it is important to point out that the present study has not
attempted to settle the key question of the appropriate inner products
for the covariant wavefunctions. Such further structure is under study,
and can be expected to be important for the realisation of modular
invariance in the models. Further applications, such as the
identification of the correct supermultiplets to which the scalar and
Dirac propagators belong, will provide the rudiments of a theory of
quantised fields at the $osp(d,2/2)$ level.

\appendix
\renewcommand{\thesection}{\Alph{section}}
\setcounter{section}{0}
\setcounter{equation}{0}
\renewcommand{\theequation}{\thesection.\arabic{equation}}
\vspace{1cm}

\app{$(a,b)$ representation of physical states}
\label{realis}

\subsection{Preliminary construction}

We can define a Heisenberg-like algebra as follows
\beq
\begin{array}{ccc}
\left[ a,b^\dagger\right] & = 1 = & \left[ b,a^\dagger \right], \\
 \left[ a,b \right] & = 0 = & \left[ b,a \right], \\
 \left[ a,a^\dagger \right] & = 0 = & \left[ b,b^\dagger \right], \\
\end{array}
\eeq
where we can take $a|0,0\rangle = 0 = b|0,0\rangle,$ and define
\beq
|m,n\rangle = (\adag)^m(\bdag)^n|0\rangle, \;\; m,n \geq 0,
\eeq
note that this implies
\bea
\langle 0,1|1,0\rangle = \langle 0,0|\bdag a | 0,0 \rangle & =&
1,\nonumber \\
\langle 1,0 | 0,1 \rangle & = & 1.
\eea
In fact, in general we have
\beq
\langle m',n'|m,n\rangle = m! n! \delta_{m'n}\delta_{n'm},
\eeq
and so we redefine our basis by
\bea
|m,n\rangle' & = & a^{\dagger m} b^{\dagger n} |0,0\rangle, \nonumber\\
|m,n\rangle &=& \frac{1}{\sqrt{m!n!}}|m,n\rangle' = \frac{(\adag)^m (\bdag)^n}{\sqrt{m!n!}} |0,0\rangle,
\label{eq:mn}
\eea

\subsection{Realisation of $\zeta_\alpha, \hat{\zeta}_\alpha$}

As explained in Section \ref{iosp}, the operators $\zeta_\alpha,
\tilde{\zeta}_\alpha$ are constructed using a two dimensional Bosonic
oscillator algebra $(a,b)$. We choose to define $\zeta_\alpha,
\tilde{\zeta}_\alpha$ as follows
\bea
\zeta_\alpha &=& \frac{1}{\sqrt{2}} \left( (ib\pm a) - (i\bdag \mp
\adag) \right), \nonumber \\
\tilde{\zeta}_\alpha &=& \frac{1}{\sqrt{2}} \left( (ia \pm b) - (i\adag
\mp \bdag) \right).
\label{eq:eigen}
\eea
At the same time we define the ghost state $\eta^\alpha$ and its
conjugate momentum $\rho_\alpha$ as
\beq
\begin{array}{lcrlcr}
\eta^1 &=& \frac{i}{\sqrt{2}} (a - \adag), & \eta^2 &=&
\frac{i}{\sqrt{2}} (b - \bdag), \\
\rho_1 &=& \frac{1}{\sqrt{2}} (b + \bdag), & \rho_2
&=&\frac{1}{\sqrt{2}} (a + \adag),
\end{array}
\eeq
which leads to
\bea
(\eta\cdot\zeta) &=& i(b-\bdag) = \sqrt{2} \eta_2, \nonumber\\
(\eta'\cdot\zeta) &=& - (a+ \adag) = -\sqrt{2} \rho_2.
\eea
In Section \ref{physstates} the eigenstates of
$(\eta\cdot\zeta)(\eta'\cdot\zeta)$ , with eigenvalues $\ell$ were
required in the analysis of the physical states. From equation
\eqn{eq:eigen} we can write
\beq
(\eta\cdot\zeta)(\eta'\cdot\zeta) = i(\adag\bdag -ab + a\bdag - \adag
b).
\eeq
We have
\beq
\adag \bdag | m,n \rangle = \sqrt{(m+1)(n+1)}| m+1,n+1 \rangle,
\eeq
and
\bea
ab|m,n\rangle &=&\frac{b^{\dagger m} a b^{\dagger n}}{\sqrt{m!n!}}|m+1,n+1\rangle,
\nonumber\\
&=&\frac{[b,a^{\dagger m}][a,b^{\dagger n}]}{\sqrt{m!n!}} |m,n\rangle,
\nonumber\\
&=&\sqrt{mn}|m-1,n-1\rangle.
\eea
Similarly
\bea
\adag b | m,n\rangle &=&
\frac{\adag[b,a^{\dagger m}]a^{\dagger n}}{\sqrt{m!n!}}|m,n\rangle =
m|m,n\rangle, \\
a\bdag | m,n\rangle &=&
\frac{a^{\dagger m}[a,b^{\dagger (n+1)}]}{\sqrt{m!n!}}|m,n\rangle = (n+1)|m,n\rangle.
\eea

As $(\eta\cdot\zeta)(\eta'\cdot\zeta)$ commutes with $(\adag b - \bdag a)$ (the false ghost number), we specialise to eigenstates $|\Lambda\rangle$ with $m=n$
\beq
|\Lambda\rangle = \sum_0^\infty \Lambda_m |m,m\rangle,
\eeq
therefore
\bea
(\eta\cdot\zeta)(\eta'\cdot\zeta)|\Lambda \rangle &=& i(\adag\bdag -ab + a\bdag - \adag b) | \Lambda \rangle, \nonumber \\
 &=& i \sum_{m=0}^\infty \left[(m+1)\Lambda_m|m+1,m+1\rangle - \Lambda_m |m,m\rangle - m\Lambda_m|m-1,m-1\rangle\right], \nonumber\\
&=&i \left[ \Lambda_0|1,1\rangle -\Lambda_0 |0,0\rangle + 2\Lambda_1|2,2\rangle - \Lambda_1 | 1,1\rangle -\Lambda_1 |0,0\rangle \right. \nonumber \\
&&\left. + 3\Lambda_2 |3,3\rangle - \Lambda_2 |2,2\rangle - 2\Lambda_2|1,1\rangle + \ldots \right], \nonumber\\
&=& i\left[-(\Lambda_0 + \Lambda_1)|0,0\rangle + (\Lambda_0 - \Lambda_1 -2\Lambda_2)|1,1\rangle +
(2\Lambda_1 -\Lambda_2 - 3\Lambda_3)|2,2\rangle + \right. \nonumber\\
&&  \left. \ldots + \left(m\Lambda_{m-1} - \Lambda_m - (m+1) \Lambda_{m+1}
\right)|m,m\rangle + \ldots\right], \nonumber\\
&=& \Lambda(\Lambda_0|0,0\rangle + \Lambda_1|1,1\rangle + \ldots +
\Lambda_m|m,m\rangle+ \ldots. \nonumber
\eea
And so
\bea
-i (\Lambda_0 + \Lambda_1) &=& \Lambda \Lambda_0, \nonumber\\
i(\Lambda_0 -\Lambda_1 - 2\Lambda_2) & =& \Lambda \Lambda_1, \nonumber\\
i(2\Lambda_1 - \Lambda_2 - 3 \Lambda_3) &=& \Lambda \Lambda_2, \nonumber
\eea
or in general
\beq
i(m\Lambda_{m-1} - \Lambda_m - (m+1) \Lambda_{m+1}) = \Lambda \Lambda_m.
\eeq
Re-expressing these in terms of $\Lambda$ and $\Lambda_0$ only we get
\bea
\Lambda_1 & = & (i\Lambda -1)\Lambda_0, \nonumber\\
\Lambda_2 &=& -\half(\Lambda^2 +2i\Lambda -2) \Lambda_0, \nonumber\\
\Lambda_3 &=& \frac{1}{6} (- i\Lambda^3 + 3\Lambda^2 + 8i\Lambda -6) \Lambda_0, \label{eq:lambdaseries}\\
\Lambda_4 &=& \frac{1}{24}(\Lambda^4 + 4i\Lambda^3 - 20\Lambda^2 -32 i \Lambda +24) \Lambda_0,
\nonumber\\
\Lambda_5 &=& \frac{i}{120}(\Lambda^5 + 5 i \Lambda^4 -40 \Lambda^3 - 100 i \Lambda^2 + 184 \Lambda + 120 i)\Lambda_0. \nonumber\\
&{\mbox etc.}& \nonumber
\eea
It is easy to write a short program to generate $\Lambda_m$ to any
order.

\subsection*{Acknowledgements}
The authors would like to thank Jan Govaerts
for continuous encouragement and constructive suggestions and feedback on
aspects
of this work. The authors also thank Alex
Kalloniatis and Peter West for discussions. Part of this work was
gestated while one of the authors
(PDJ) was on study leave at NIKHEF, Amsterdam, and the hospitality of Jan
Willem van Holten
and the NIKHEF theory group is acknowledged. PDJ further acknowledges
financial support from the Alexander von Humboldt Foundation. Finally, IT
acknowledges the
Australian Research Council for the award of a Fellowship, and Tony Bracken
 and the Centre for Mathematical Physics, University of Queensland, for 
support. Finally SC acknowledges support from an Australian Postgraduate Award.


\begin{thebibliography}{10}

\bibitem{scalar}
Jarvis~P D and Tsohantjis I.
\newblock {\em J. Phys. A: Math. Gen.}, 29:1245, 1996.

\bibitem{wigsym}
Jarvis~P D, Bracken~A J, Corney~S P, and Tsohantjis I.
\newblock Realisations of physical particle states via cohomolgies:
  algebraization of brst-bfv covariant quantization.
\newblock In P~Kasperkovitz and D~Grau, editors, {\em 5th Wigner Symposium},
  Singapore, 1998. World Scientific.

\bibitem{clausthal}
Jarvis~P D, Bracken A, Corney~S P, and Tsohantjis I.
\newblock Lie super-algebraisation of brst-bfv covariant quantisation.
\newblock In H-D Doebner, V~K Dobrev, and J~Hilgert, editors, {\em Lie Theory
  and is Applications in Physics II}, Singapore, 1998. World Scientific.

\bibitem{d21agrp22}
Corney~S P, Jarvis~P D, and McAnally~D S.
\newblock The d(2,1/a) particle.
\newblock In Corney~S P, Delbourgo R, and Jarvis~P D, editors, {\em Int. Coll.
  on Group Theor. Methods in Physics}, Boston, 1999. International Press.

\bibitem{casalbuoni}
A~Barducci, R~Casalbuoni, D~Dominici, and R~Gatto.
\newblock {\em Phys. Lett. B}, 187:135, 1987.

\bibitem{brink}
Brink L., Deser S., Zumino B, Di~Vecchia P, and Howe P.
\newblock {\em Phys. Lett. B}, 64:435, 1976.

\bibitem{brinkschwarz}
Brink L and Schwarz~J H.
\newblock {\em Phys. Lett. B}, 100:310, 1981.

\bibitem{brinkgreen}
Brink L and Green~M B.
\newblock {\em Phys. Lett. B}, 106:393, 1981.

\bibitem{govbk}
J~Govaerts.
\newblock {\em Hamiltonian quantisation and constrained dynamics}.
\newblock Leuven University Press, Belgium, 1991.

\bibitem{pdjhsg}
Jarvis P and Green~H S.
\newblock {\em J. Math. Phys}, 20:2115, 1979.

\bibitem{hartley}
Hartley~D H and Cornwell~J F.
\newblock {\em J. Phys. A: Math. Gen.}, 21:3171, 1988.

\bibitem{macksalam}
Mack G and Salam A.
\newblock {\em Annals of Physics}, 53:174, 1969.

\bibitem{henneaux}
M~Henneaux.
\newblock {\em Phys. Lett. B}, 177:35, 1986.

\bibitem{BFV1}
G~A~Vilkovisky E~S~Fradkin.
\newblock {\em Phys. Lett. B}, 55:224, 1975.

\bibitem{BFV2}
G~A~Vilkovisky J~A~Batalin.
\newblock {\em Phys. Lett. B}, 69:309, 1977.

\bibitem{BFV3}
Fradkin~F S and Fradkina~T E.
\newblock {\em Phys. Lett. B}, 72:343, 1978.

\bibitem{monaghan}
S~Monaghan.
\newblock {\em Phys. Lett. B}, 178:231, 1986.

\bibitem{teitelboim1}
M~Henneaux \&~C Teiltelboim.
\newblock {\em Phys. Rev. D}, 143:127, 1982.

\bibitem{teitelboim2}
C~Teitelboim.
\newblock {\em Phys. Rev. D}, 25:3159, 1982.

\bibitem{teitelbk}
M~Henneaux~C Teiltelboim.
\newblock {\em Quantisation of gauge systems}.
\newblock Princeton University Press, Princeton, 1992.

\bibitem{hwang}
Hwang S.
\newblock {\em Nucl. Phys. B}, 386:231, 1984.

\bibitem{kappasusy}
Jarvis~P D, van Holten J~W, and Kowalski-Glikman J.
\newblock {\em Phys. Lett. B}, 427:47, 1998.

\bibitem{maldacena}
Maldacena~J M.
\newblock {\em Adv. Theor. Math. Phys.}, 2:231, 1998.

\bibitem{witten}
Witten E.
\newblock {\em Adv. Theor. Math. Phys.}, 2:253, 1998.

\bibitem{siegel}
Siegel W and Zweibach B.
\newblock {\em Nucl. Phys. B}, 282:125, 1987.

\bibitem{bars1}
Bars A, Deliduman C, and Andreev O.
\newblock {\em Phys. Rev. D}, 58:066004, 1998.

\bibitem{bars2}
Bars I.
\newblock Two-time physics.
\newblock In Corney~S P, Delbourgo R, and Jarvis~P D, editors, {\em Int. Coll.
  on Group Theor. Methods in Physics}, Boston, 1999. International Press.

\bibitem{bars3}
Bars I, Deliduman C, and Minic D.
\newblock {\em Phys. Rev. D}, 59:125004, 1998.

\bibitem{bracken1}
Bracken~A J and Jessup B.
\newblock {\em J. Math. Phys}, 23:1925, 1982.

\bibitem{bracken2}
Bracken~A J and Jessup B.
\newblock {\em J. Math. Phys}, 23:1947, 1982.

\end{thebibliography}
\end{document}